\documentclass[a4paper,11pt]{article}
\usepackage{jheppub}
\usepackage{lineno}
\usepackage{cleveref}

\title{Vacuum, ma non troppo:\\
hidden matter distribution in symmetry-transformed electrovacuum spacetimes}

\author{C. Herdeiro and J. Novo}
\affiliation{Departamento de Matem\'atica da Universidade de Aveiro and
    Center for Research and Development in Mathematics and Applications (CIDMA),\\
    Campus de Santiago, 3810-193 Aveiro, Portugal}
\emailAdd{herdeiro@ua.pt,j.novo@ua.pt}

\abstract{
We analyse two static spacetimes, recently generated from a Schwarzschild--Bertotti--Robinson electrovacuum seed through distinct symmetry transformations. The electromagnetic field of the transformed solutions can be set to zero and both solutions were presented as vacuum solutions of General Relativity.
However, we show that once recast in Weyl form, both metrics are seen to be supported by a semi-infinite annular mass distribution on the equatorial plane.
Thus, these metrics harbour a hidden matter source, visible in Weyl coordinates.
In the original Schwarzschild-like coordinates equatorial null geodesics reach spatial infinity ($r\to\infty$) in finite affine parameter. In the Weyl representation, these geodesics approach the inner edge of the annulus, where the coordinate map degenerates.
}

\numberwithin{equation}{section}
\keywords{black holes, exact solutions, singularities, Weyl metrics, solution-generating techniques}

\begin{document}
\maketitle
\flushbottom

\section{Introduction}

The classification and physical interpretation of exact solutions to Einstein's equations remain a central endeavour in General Relativity.
While the Kerr--Newman family exhaustively describes stationary black holes in electrovacuum (with a connected event horizon) under standard asymptotic conditions~\cite{Robinson:1975bv,Chrusciel:2012jk}, relaxing these assumptions, for instance with extra fields or extra couplings, opens the door to a rich landscape of hairy black-hole solutions - see e.g.~\cite{Herdeiro:2015waa}.
Understanding which of these may represent physically viable spacetimes requires, to start with, careful analysis of their global structure, geodesic completeness, and asymptotic behaviour, besides critical issues such as stability.

A useful strategy for constructing new families is to start from an electrovacuum ``seed'' and apply solution-generating symmetries (e.g. of Ernst type), thereby producing non-trivial geometries. Electromagnetic backgrounds of Bertotti--Robinson and Bonnor--Melvin type~\cite{Bertotti:1959pf,Robinson:1959ev,Bonnor:1954tis,Melvin:1963qx} provide a natural arena in which to apply these symmetry transformations.
In certain limits of the so generated solutions, the electromagnetic field can be removed yielding, apparently, vacuum solutions. However, as we point out in this note, the removed field can manifest as an effective matter distribution, which may be hidden in some coordinate patches, but visible in the maximal analytic extension of the geometries.

Two concrete examples are discussed here to underscore this point.
The first is the static ``hairy" black hole of~\cite{Astorino:2026okd}, obtained from a Schwarzschild--Bertotti-Robinson (SBR) seed by magnetisation and the subsequent removal of the electromagnetic (EM) field by an appropriate choice of parameter.
The second, constructed explicitly in~\cite{Barrientos:2026shy}, uses azimuthal inversion and yields a Schwarzschild--Levi-Civita-like branch~\cite{Barrientos:2024uuq,Barrientos:2025rjn}. At first sight these families appear independent,  but connections between them have been suggested in the literature. Here, we unveil a common underlying structure.

First, we show that in the original (Schwarzschild-like) coordinates, equatorial null geodesics reach spatial infinity, $r\to \infty$, in finite affine parameter. This geodesic incompleteness persists whether a black hole horizon is present or not. We therefore analyse the horizonless case and then argue the same conclusion for the horizonful geometry.

Then, we consider the canonical Weyl form for both solutions~\cite{Weyl:1917gp,Bach:1922lav,Stephani:2003tm}. The map between the original coordinates and the Bach--Weyl coordinates is the same for both solutions. Studying the corresponding Weyl forms, we show that both metrics share the same geometric support: a singular semi-infinite annular mass distribution on the equatorial plane.
The only difference is the sign of the effective surface density, and the nature of the rod.
Critically, in the original coordinates, equatorial null geodesics leave the coordinate patch in finite affine parameter. In the Weyl representation these geodesics terminate precisely at the inner edge of this annulus in finite affine parameter. Thus, despite being generated from electrovacuum seeds, the transformed spacetimes are not truly vacuum; their Ricci tensor has a distributional component compatible with an effective distributional source supported on a semi-infinite annulus. This structure is not evident in the original Schwarzschild-like coordinates.

These results reveal a non-trivial structural link between the two branches. As noted in~\cite{Astorino:2026xwz}, 
a singular rescaling of coordinates and parameters can relate the local form of the magnetised and inversion-generated branches. This relation, however, does not imply global equivalence of the corresponding spacetimes. Instead, it points to a deeper structural connection, which in our analysis is made precise by the fact that both solutions share the same Weyl map and are supported by the same annular distribution, differing only in the sign of the effective density.

The paper is organised as follows.
In \cref{sec:seed} we summarise the common SBR seed and the two generating maps.
In \cref{sec:A} and \cref{sec:B} we analyse, respectively, the magnetised and inversion branches (metric form, geodesic completion, and Weyl-source interpretation).
In \cref{sec:relation} we discuss the relation between the two families in light of the symmetry analysis in~\cite{Astorino:2026xwz}.
Final remarks are presented in \cref{sec:conclusions}.

\section{Common SBR seed and symmetry maps}
\label{sec:seed}
We take as seed the static SBR electrovacuum configuration in spherical-type coordinates:
\begin{equation}
\mathrm{d}s^2 = \frac{1}{\Omega^2}\left[-Q\mathrm{d}t^2 + \frac{\mathrm{d}r^2}{Q} + r^2\left(\frac{\mathrm{d}\theta^2}{P} + P\sin^2\theta\,\mathrm{d}\phi^2\right)\right],\label{eq:seedSBR}
\end{equation}
with
\begin{align}
P &= 1 + B^2 m^2\cos^2\theta, \\
Q &= \left(1-\frac{2m}{r}-B^2m^2\right)\left(1+B^2r^2\right), \\
\Omega &= \sqrt{1+B^2r\left[r+\left(2m+B^2m^2r-r\right)\cos^2\theta\right]},
\end{align}
and magnetic potential
\begin{equation}
A = \frac{1}{B}\left(r\partial_r\Omega-\Omega+1\right)\mathrm{d}\phi.
\end{equation}
The parameters are the mass scale $m$ and the external-field scale $B$.

The solution possesses a conical singularity on the rotation axis with strength:
\begin{equation}
\lim_{\theta\to \{0,\pi\}}\frac{1}{\sin\theta}\intop_0^{2\pi}\frac{\sqrt{g_{\phi\phi}}}{\sqrt{g_{\theta\theta}}}=2\pi\left(1+B^2 m^2\right)\,.
\end{equation}
This can be removed by tuning the periodicity of $\phi$.

Starting from this same seed, one can construct two vacuum-like branches:
\begin{enumerate}
\item a Harrison-type magnetisation followed by parameter tuning, giving the static hairy family of~\cite{Astorino:2026okd};
\item an azimuthal inversion map as in~\cite{Barrientos:2026shy}, yielding a Schwarzschild--Levi-Civita-type branch.
\end{enumerate}

Both procedures remove the physical electromagnetic field in the final configuration while retaining a non-trivial metric deformation inherited from the electrovacuum seed. In the next two sections we consider both cases in turn.

\section{Case A: Magnetised branch}
\label{sec:A}

\subsection{Metric and background}
The hairy black hole solution of~\cite{Astorino:2026okd} can be obtained by immersing the seed metric of \cref{eq:seedSBR} into an external Bonnor--Melvin EM field. It is possible to tune the strengths of the fields such that the magnetic potential vanishes. The resulting metric can be written as:
\begin{align}
\mathrm{d}s^2 &= \frac{\left(1 + B^2 m r \cos^2\theta + \Omega\right)^2}{4\Omega^4}\left[-Q\,\mathrm{d}t^2+\frac{\mathrm{d}r^2}{Q}+\frac{r^2}{P}\mathrm{d}\theta^2\right] \nonumber \\
&\quad +\frac{4r^2P\sin^2\theta}{\left(1 + B^2 m r \cos^2\theta + \Omega\right)^2}\,\Delta_\phi^2\,\mathrm{d}\phi^2.
\label{eq:metricA}
\end{align}
With $\Delta_\phi=(1+B^2m^2)^{-1}$, the axis is free of conical singularities.
The horizon is located at
\begin{equation}
r_H = \frac{2m}{1-B^2m^2},
\end{equation}
which is well-defined and can be interpreted as a standard horizon when $m>0$ and $|Bm|<1$.

The $m=0$ background is
\begin{equation}
\mathrm{d}s^2 = \frac{(1 + \Omega)^2}{4\Omega^4}\left[-(1+B^2r^2)\mathrm{d}t^2 + \frac{\mathrm{d}r^2}{1+B^2r^2} + r^2\mathrm{d}\theta^2\right]
+ \frac{4r^2\sin^2\theta}{(1+\Omega)^2}\mathrm{d}\phi^2,
\label{eq:bgA}
\end{equation}
with $\Omega=\sqrt{1+B^2r^2\sin^2\theta}$. This geometry is Ricci flat and its Kretschmann scalar, given by
\begin{equation}
    R_{\mu\nu\alpha\beta}R^{\mu\nu\alpha\beta} = 768B^4\frac{ \left( 1 + B^2r^2\sin^2\theta \right)^3 }{\left( 1+ \sqrt{1+B^2 r^2 \sin^2 \theta}\right)^6}\,,
\end{equation}
is regular everywhere.

\subsection{Geodesic incompleteness}
For equatorial radial null geodesics ($\theta=\pi/2$), one finds
\begin{equation}
\frac{\mathrm{d}\lambda}{\mathrm{d}r}=\pm\frac{\sqrt{-g_{tt}g_{rr}}}{E}
=\pm\frac{\left(1+\sqrt{1+B^2r^2}\right)^2}{4(1+B^2r^2)^2}\frac{1}{E},
\end{equation}
which integrates to
\begin{equation}
\Delta\lambda
=\frac{1}{E}\int^{\infty}\frac{\left(1+\sqrt{1+B^2r^2}\right)^2}{4(1+B^2r^2)^2}\,\mathrm{d}r
<\infty.
\label{eq:afinA}
\end{equation}
Here, $E$ is the Noether charge associated with the Killing vector field $\partial_t$. Since the asymptotic structure of the spacetime is not clear, we do not strictly identify it with particle energy, but keep the notation $E$ for familiarity.
Taking as an example a particle emitted at the origin this yields:
\begin{equation}
\Delta\lambda =\frac{8+3\pi}{16|B|E}\,.
\end{equation}
Hence photons reach $r\to\infty$ in finite affine parameter.
Because $\sqrt{-g_{tt}g_{rr}}$ on the equator is insensitive to the $Q$ factor in \cref{eq:metricA}, and because the equatorial restriction removes the $m$ term in $\Omega$, the same incompleteness persists in the full $m\neq0$ geometry along equatorial radial null geodesics.

This indicates that either the geodesics terminate at an irremovable singularity, or that the Schwarzschild-like chart does not provide a maximal development of null geodesics\footnote{The ``point'' reached ($r\rightarrow \infty$ and $\theta=\pi/2$) is non-singular in this patch.}. This will be addressed in the next subsection.

\subsection{Weyl map and source support}
Since the presence of the horizon does not alter the fact that null geodesics can leave the coordinate patch we focus on the horizonless case, $m=0$ to analyse the asymptotic structure of the spacetime.  Writing \cref{eq:bgA} in Weyl form,
\begin{equation}
\mathrm{d}s^2 = -e^{2U}\mathrm{d}t^2 + e^{-2U}\left[e^{2K}(\mathrm{d}\rho^2+\mathrm{d}z^2)+\rho^2\mathrm{d}\phi^2\right],
\end{equation}
comparison with~\eqref{eq:bgA} yields
\begin{align}
    \label{eq:rhoeqA} \rho^2 = (1+B^2 r^2)\frac{r^2\sin^2\theta}{\Omega^4}\,,\quad
    \mathrm{e}^{2U} = (1+B^2 r^2)\frac{(1+\Omega)^2}{4\Omega^4}\,.
\end{align}

The Weyl coordinates $(\rho, z)$ are constructed by requiring the $(\rho, z)$ part of the metric to be conformally flat. A first step in doing so is to introduce an isothermal radial coordinate $R$ through
\begin{equation}
\mathrm{d}R=\frac{\mathrm{d}r}{r\sqrt{1+B^2r^2}}.
\end{equation}
The $(R,\theta)$ sector is conformally flat, so the map $(R,\theta)\mapsto(\rho,z)$ must satisfy Cauchy--Riemann equations,
\begin{equation}
\frac{\partial z}{\partial R}=\frac{\partial \rho}{\partial\theta},
\qquad
\frac{\partial z}{\partial\theta}=-\frac{\partial \rho}{\partial R}.
\end{equation}
Using the chain rule with
\begin{equation}
W(r)=\sqrt{\frac{1}{r^2(1+B^2r^2)}},
\end{equation}
these become
\begin{equation}
\frac{\partial z}{\partial r}=W(r)\frac{\partial \rho}{\partial\theta},
\qquad
\frac{\partial z}{\partial\theta}=-\frac{1}{W(r)}\frac{\partial \rho}{\partial r}.
\end{equation}
Substituting $\rho^2=(1+B^2r^2)r^2\sin^2\theta/\Omega^4$ yields
\begin{equation}
\frac{\partial z}{\partial r}=\frac{\cos\theta\left(1 - B^2 r^2 \sin^2\theta\right)}{\Omega^4},
\qquad
\frac{\partial z}{\partial\theta}=-\frac{r\left[2(1+B^2r^2)-\Omega^2\right]\sin\theta}{\Omega^4},
\end{equation}
whose integrable solution is precisely
\begin{equation}
z=\frac{r\cos\theta}{\Omega^2}=\frac{r\cos\theta}{1+B^2r^2\sin^2\theta}.\label{eq:zeqA}
\end{equation}
This, together with \cref{eq:rhoeqA}, specifies the Weyl map.

\begin{figure}
    \centering
    \includegraphics[width=0.8\linewidth]{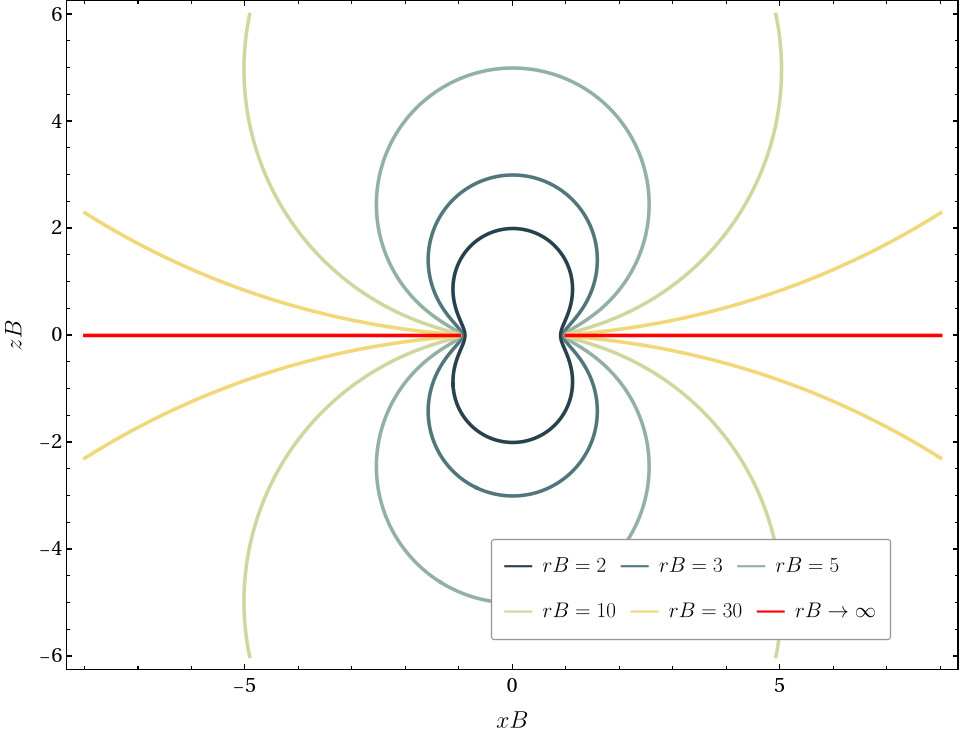}
    \caption{Lines of constant $rB$ in $y=0$ cross section of the Euclidean $\mathbb{R}^3$, where $x=\rho\sin\theta$, $y=\rho\cos\theta$, and $z=z$ (this is not an isometric embedding). The limit $rB\to\infty$ corresponds to a semi-infinite annulus on the equatorial plane starting at $\rho B=1$, this region is highlighted in red. All other curves are closed.}
    \label{fig:constant_r}
\end{figure}

The determinant of the Jacobian, $\mathcal{J}$, of this transformation reads:
\begin{equation}
    \det(\mathcal{J})=-r\frac{1+B^2r^2\cos^2 \theta}{\sqrt{1+B^2r^2} \left(1+B^2r^2 \sin^2\theta\right)^2}\,.
\end{equation}
This is everywhere finite with the exception of the limit $r\to\infty$, where the determinant vanishes, and one has:
\begin{equation}
    \rho=\frac{1}{|B|\sin\theta}\,,\quad z=0.
\end{equation}
When varying $\theta$ this corresponds to the semi-infinite line $\rho>1/|B|\,,z=0$ on the $(\rho,z)$ plane. This generates a semi-infinite annulus when translated along the spacelike Killing vector field $\partial_\phi$. This is also illustrated in \Cref{fig:constant_r}, where several lines of constant $r$ are plotted in the Cartesian-like $(x=\rho\sin\theta,z)$ plane (note that this is merely for illustrative purposes, and is not an isometric embedding). These lines were obtained using the following relation:
\begin{equation}
    \frac{1}{B^2}\left(\frac{1}{\rho ^2+z^2}-\frac{1}{r^2}\right)+\frac{r^2 z^2}{\left(\rho ^2+z^2\right)^2}=1\,.
\end{equation}

The fact that the transformation between the original Schwarzschild-like coordinates and the Weyl ones is not invertible at $r\to \infty$ seems to point to some underlying structure of the spacetime. Recall that the Weyl map for the Schwarzschild coordinates is also not invertible at the horizon's location. In fact, it will be shown below that the annulus where the Weyl map is not invertible can be interpreted as a distributional source.

In the Bach--Weyl formalism, the potential $U$ satisfies the flat-space Laplace equation $\nabla^2_{\mathbb{E}^3} U = 0$ away from possible sources, corresponding to sets of zero measure, allowing interpretation as a Newtonian gravitational potential~\cite{Stephani:2003tm}. The explicit form of this potential in $(\rho,z)$ coordinates is too convoluted and provides no simple interpretation. Instead, as the possible source is located at an annulus we resort to oblate spheroidal coordinates, given by
\begin{equation}
\rho=a\cosh\mu\sin\nu,
\qquad
z=a\sinh\mu\cos\nu,
\end{equation}
with $\mu \geq 0$, $0 < \nu < \pi$, and $a = 1/|B|$\footnote{Note that the above transformation is not invertible when $\nu=\pi/2$ corresponding precisely to the annulus $\rho>1/|B|\,,z=0$, nonetheless these are the coordinates best adapted to the symmetries of the problem.}, such coordinates have been previously used to describe black holes surrounded or not with general relativistic disks~\cite{Morgan:1969jr,Lemos:1993uy}. 
Then, the potential takes the rather simple form:
\begin{equation}
U(\mu,\nu)=\log\!\left[\frac{\cosh\mu}{2}(1+|\cos\nu|)\right].\label{eq:potA}
\end{equation}

The appearance of $|\cos\nu|$ signals trouble at $\nu = \pi/2$ (the equatorial plane $z = 0$). While $U$ itself is continuous there, its gradient is not:
\begin{equation}
    \left.\partial_\nu U\right|_{\nu = \pi/2^{\pm}} = \pm 1\,.
\end{equation}
This discontinuity indicates a surface mass distribution in the Weyl sense. The surface $\nu = \pi/2$ corresponds to
\begin{equation}
    \rho = \frac{\cosh\mu}{|B|}\,, \quad z = 0\,,
\end{equation}
which describes a semi-infinite annulus: the region $\rho \geq 1/|B|$ in the equatorial plane, or equivalently, an infinite plane with a circular hole of radius $1/|B|$ centred at the origin.

Applying Gauss's law for gravity, where the gravitational field is related to the Weyl potential by $\mathbf{g}=-\nabla U$, the surface mass density is defined as
\begin{equation}
    \sigma = -\frac{1}{4\pi}(\mathbf{g}^+ - \mathbf{g}^-)\cdot\mathbf{n}\,,
\end{equation}
where $\mathbf{n}$ is the upward unit normal and $\mathbf{g}^{\pm}$ denotes the gravitational field when approaching from above/below. For the metric (\ref{eq:bgA}) this yields:
\begin{equation}
    \sigma(\rho) = - \frac{1}{4\pi} \left. \frac{\partial U}{\partial z}\right|^{z=0^+}_{z=0^-} =  \frac{|B|}{2\pi \sqrt{B^2\rho^2 - 1}}\,, \quad \rho > \frac{1}{|B|}\label{eq:sigmaA}\,. 
\end{equation}
This density diverges at the inner edge $\rho = 1/|B|$ and falls off as $1/\rho$ at large distances.

Consequently, the metric (\ref{eq:bgA}) is only locally Ricci flat, $R_{\mu\nu}=0$. The distributional source supported on the equatorial plane $z=0$ for $\rho>1/|B|$ leads to delta-function contributions in the curvature. In particular, the Ricci tensor acquires a distributional part of the form $R_{\mu\nu}^{\rm dist}=\mathcal{R}_{\mu\nu}(\rho)\,\delta(z)$, whose non-vanishing orthonormal components read
\begin{align}
\mathcal{R}_{t}^{t} &= -8 \left|B\right|\sqrt{B^2\rho^2-1}\,\Theta(\rho-1/|B|)\,\\
\mathcal{R}_{\rho}^{\rho} &= \mathcal{R}_{z}^{z} = -8 \left|B\right|\sqrt{B^2\rho^2-1}\,\Theta(\rho-1/|B|)\,\\
\mathcal{R}_{\phi}^{\phi} &= 8 \left|B\right|\sqrt{B^2\rho^2-1}\,\Theta(\rho-1/|B|)\,.
\end{align}
These terms encode the same semi-infinite annular mass distribution identified from the jump of the potential, cf. \cref{eq:sigmaA}, and show explicitly that the spacetime, while vacuum in a local sense, is globally supported by a distributional source. Interestingly, none of these terms diverge and are all $\mathcal{C}^0$, when restricted to $z=0$.

\section{Case B: Inversion-generated branch}
\label{sec:B}

\subsection{Metric and background}
Starting again with \cref{eq:seedSBR}, one can apply a symmetry transformation that yields a spacetime whose EM field is purely gauge, and which therefore satisfies Einstein's vacuum equations. This was done in~\cite{Barrientos:2026shy}, yielding the metric:
\begin{equation}
\mathrm{d}s^2=\frac{r^2P\sin^2\theta}{f\Omega^4}
\left[-Q\mathrm{d}t^2+\frac{\mathrm{d}r^2}{Q}+\frac{r^2}{P}\mathrm{d}\theta^2\right]
+f\mathrm{d}\phi^2,
\label{eq:metricB}
\end{equation}
with
\begin{equation}
f=-\frac{B^2}{4}\frac{1+B^2mr\cos^2\theta+\Omega}{1+B^2mr\cos^2\theta-\Omega}.
\end{equation}
Its $m=0$ background is
\begin{equation}
\mathrm{d}s^2=\frac{4r^2\sin^2\theta}{B^2\Omega^4}\frac{\Omega-1}{\Omega+1}
\left[-(1+B^2r^2)\mathrm{d}t^2+\frac{\mathrm{d}r^2}{1+B^2r^2}+r^2\mathrm{d}\theta^2\right]
+\frac{B^2}{4}\frac{1+\Omega}{1-\Omega}\mathrm{d}\phi^2,
\label{eq:bgB}
\end{equation}
again with $\Omega=\sqrt{1+B^2r^2\sin^2\theta}$.

When $B=0$ and $m=0$, this solution reduces to the Levi--Civita spacetime~\cite{Stephani:2003tm}, although the $B\to0$ limit must be taken carefully. It can therefore be viewed as a one-parameter extension of the Schwarzschild--Levi--Civita spacetime~\cite{Akbar:2015qna}. The Levi--Civita character of \cref{eq:metricB} manifests itself in the absence of a rotation axis, since $g_{\phi\phi}$ never vanishes and there are no fixed points of $\partial_\phi$. Moreover, the metric is singular at the expected location of this axis, $\sin\theta=0$, as well as  at $r=0$. These singularities are naked when not covered by a horizon. These singularities can be seen by considering the Kretschmann scalar
\begin{equation}
    R_{\mu\nu\alpha\beta}R^{\mu\nu\alpha\beta} = 3 B^{12}\frac{ \left( 1 + B^2r^2\sin^2\theta \right)^3 }{\left( 1-\sqrt{1+B^2 r^2 \sin^2 \theta}\right)^6}\,.
\end{equation}
Due to the minus sign on the denominator the Kretschmann scalar diverges whenever $r \sin\theta=0$.

\subsection{Geodesic incompleteness}
Along equatorial radial null geodesics,
\begin{equation}
\frac{\mathrm{d}\lambda}{\mathrm{d}r}
=\pm\frac{4\left(1-\sqrt{1+B^2r^2}\right)^2}{B^4(1+B^2r^2)^2E},
\end{equation}
so
\begin{equation}
\Delta\lambda=\frac{1}{E}\int^{\infty}\sqrt{-g_{tt}g_{rr}}\,\mathrm{d}r<\infty,
\end{equation}
for example $\Delta\lambda=(3\pi-8)/(E\left|B\right|^5)$ for a trajectory starting from $r=0$.
Therefore, as with Case A, geodesics terminate in finite affine parameter.

As in the magnetised branch, the finite affine distance implies that $r\to\infty$ is not a benign asymptotic region of the physical manifold.
The coordinate patch ends at finite affine parameter, so the corresponding spacetime region should be viewed as a non-maximal chart. In fact, the Weyl representation discussed next shows that, on the Weyl side, the natural endpoint of these geodesics is the inner edge of a distributional annular source.

\subsection{Same Weyl map, opposite jump}
Once more, the global structure of the background, $m=0$, will be considered.
A key point from~\cite{Barrientos:2026shy} is that both transformations preserve $g_{tt}g_{\phi\phi}$ and conformally rescale the non-Killing sector in the same way.
Consequently, the Weyl map is exactly the same (given by \cref{eq:rhoeqA} and \cref{eq:zeqA}) also for \cref{eq:bgB}.

In oblate variables, the potential $U$ of \cref{eq:bgB} is
\begin{equation}
U(\mu,\nu)=\log\!\left[\frac{2}{B^2}(1-|\cos\nu|)\cosh\mu\right],
\end{equation}
so the jump reverses sign, when compared with \cref{eq:potA}, and
\begin{equation}
\sigma(\rho)=-\frac{|B|}{2\pi\sqrt{B^2\rho^2-1}},
\qquad
\rho>\frac{1}{|B|}\label{eq:sigmaB}\,.
\end{equation}
Therefore, the two branches differ by the sign of the effective Weyl surface density, while keeping the same support and singular inner edge.

Hence, the distinction between the two families lies in the sign of the matter density and in the nature of the $\rho=0$ rod: spacelike (and regular) for case A, and timelike (and singular) for case B.
Equatorial null geodesics in both cases end at the same inner edge, reinforcing that the finite-affine-parameter effect is tied to this common distributional structure.

The respective non-vanishing components of the distributional Ricci tensor are:
\begin{align}
\mathcal{R}_{t}^{t} &= \frac12 \left|B\right|^5\sqrt{B^2\rho^2-1}\,\Theta(\rho-1/|B|)\,,\\
\mathcal{R}_{\rho}^{\rho} &=\mathcal{R}_{z}^{z}=  \frac12 \left|B\right|^5\sqrt{B^2\rho^2-1}\,\Theta(\rho-1/|B|)\,,\\
\mathcal{R}_{\phi}^{\phi} &=  -\frac12 \left|B\right|^5\sqrt{B^2\rho^2-1}\,\Theta(\rho-1/|B|)\,.
\end{align}
Once again, all terms are everywhere finite, and $\mathcal{C}^0$, when restricted to the equatorial plane.

\section{Relation between the two branches}
\label{sec:relation}

The results presented above show a deep connection between the two families. Both are sourced by an annular matter distribution with the same absolute density and share the same Bach--Weyl map. This could be related by a $\mathbb{Z}^2$ inversion, were it not for the different nature of the $\rho=0$ set. Case A possesses a regular rotation axis at $\rho=0$, while case B displays a timelike $\rho=0$ rod, which is singular.

These similarities are not accidental, as the symmetry transformations leading to each branch are not independent. In fact, as shown in~\cite{Astorino:2026xwz}, one can move between them by an appropriate coordinate rescaling and a limit of the parameter $B$.

A caveat is in order. The coordinate map $(r,\theta)\to(\rho,z)$ degenerates precisely on the locus $\rho>1/|B|\,, z=0$, where the Weyl sources are located. The original chart is itself problematic there: the metric determinant vanishes and the asymptotic region $r\to\infty$ has $\mathbb{R}\times S^1$  topology\footnote{In the $r\to\infty$ limit surfaces of constant time and radial coordinate are described by the line element $\mathrm{d}s^2 = \mathrm{d}\theta^2/(4B^2\sin^2 \theta) +  4\mathrm{d}\phi^2/B^2$, for metric (\ref{eq:bgA}), and $\mathrm{d}s^2=4\mathrm{d}\theta^2/(B^6\sin^2\theta)+B^2\mathrm{d}\phi^2/4$, for metric (\ref{eq:bgB}).}   rather than the usual $S^2$ one might expect. This holds for both solutions considered here. The identification of the inner edge of the Weyl annulus with the locus reached by equatorial null geodesics in the original chart is therefore best understood as a statement about the Weyl representation of the geometry, not as a claim that the two charts are diffeomorphic in a neighbourhood of the source. What is unambiguous is that (i) equatorial null geodesics escape the original chart in finite affine parameter, and (ii) the Weyl form of the same metric carries a distributional Ricci component supported on a semi-infinite annulus; the two statements are mutually consistent and physically suggestive, but a fully rigorous gluing would require a separate analysis of the boundary structure.

\section{Conclusions}
\label{sec:conclusions}
We have reanalysed two static geometries derived from the SBR electrovacuum seed within a unified framework, emphasising a critical aspect of symmetry transformations in exact solutions.

Both solutions emerge from an electrovacuum seed, and the transformed metrics are devoid of any physical EM field, so they solve, at first sight, Einstein's equations in vacuum. However, as shown above, they are not true vacuum solutions: the Ricci tensor acquires a distributional component consistent with a semi-infinite annular mass distribution. These sources are distributional, making the (Ricci) curvature diverge as a delta function on the annulus. As such, tidal forces are not arbitrarily high, in their neighbourhood. The gravitational backreaction of the original EM field is encoded in this distribution. Interestingly, the original Schwarzschild-like coordinate patch does not include this semi-infinite annulus; it becomes apparent only after extending the metric with Bach--Weyl coordinates.

An indication of this mass distribution is already present in the original coordinates: photons can reach $r\to\infty$ in finite affine parameter, signalling it may be incomplete. In Weyl form, the cause becomes clear: the point $r\to\infty$, $\theta=\pi/2$, is mapped to a finite $\rho$, which corresponds to the inner edge of the matter distribution.

Despite carrying the classical label ``vacuum,'' these metrics are not truly vacuum in the Weyl sense.
They represent elliptic-type effective sourced geometries, a consequence of careful symmetry operations \emph{on electrovacuum} seeds without full removal of backreaction.

Thus, the two metrics are best understood as two closely related facets of a unified SBR-driven deformation mechanism, where the nature of the Weyl rod (sign of surface density) changes while the underlying annular source support and geodesic structure remain identical. An interesting future direction would be to analyse if the presence of the black hole can change the nature of these distributional sources. Although, the fact that the equatorial radial null geodesics are insensitive to the presence of the black hole suggests that at least some of the properties persist.

A conceptual parallel can be drawn between the present analysis and the zero-mass limit of the Kerr spacetime discussed in~\cite{Gibbons:2017djb}. In that case, although the metric becomes locally flat and is often described as vacuum, a distributional curvature singularity supported on a ring persists, so that the spacetime is only vacuum away from this locus. A similar phenomenon occurs here: despite the absence of an explicit matter field in the transformed solutions, the Weyl representation reveals a distributional source - now supported on a semi-infinite annulus - which is not visible in the original coordinate patch. In both settings, the apparent vacuum character is therefore only local.

The two black-hole solutions described here are part of the family of known analytical solutions in four-dimensional general relativity, as discussed in~\cite{Astorino:2026xwz}. Solutions with frame dragging are also possible; however, these cannot be interpreted in terms of a Newtonian potential, and care is likewise needed in such cases. For example, the recently presented \emph{curling} spacetime~\cite{Astorino:2026xwz} exhibits closed timelike curves, $g_{\phi\phi}<0$, close to the axis.
As another example, the recently constructed rotating Ricci-flat geometry~\cite{Ma:2026otg} obtained through demagnetisation of the Kerr–Bertotti–Robinson spacetime, may likewise conceal a distributional matter source in its maximal extension. Since its static limit reduces to the solution in~\cite{Astorino:2026okd}, analysed here, and given its similarly non-asymptotically-flat spindle structure, it is plausible that the rotating geometry may also admit a hidden annular or ring-like support, not visible in the original Boyer–Lindquist-type coordinates but revealed in a more global description.

\acknowledgments
We thank M. Astorino, A. Cisterna, P. Cunha and E. Radu for comments on this manuscript. This work is supported by the Center for Research and Development in Mathematics and Applications (CIDMA) (\url{https://ror.org/05pm2mw36}) under the Portuguese Foundation for Science and Technology 
(FCT -- Fundaç\~ao para a Ci\^encia e a Tecnologia, \url{https://ror.org/00snfqn58}), Grants UID/04106/2025 (\url{https://doi.org/10.54499/UID/04106/2025}) and UID/PRR/04106/2025 (\url{https://doi.org/10.54499/UID/PRR/04106/2025}), as well as the projects: Horizon Europe staff exchange (SE) programme HORIZON-MSCA2021-SE-01 Grant No.\ NewFunFiCO-101086251 and 2022.04560.PTDC (\url{https://doi.org/10.54499/2022.04560.PTDC}).
J.N.\ is funded by FCT through the following grant 10.54499/2021.06539.BD (\url{https://doi.org/10.54499/2021.06539.BD})

\bibliographystyle{JHEP}
\bibliography{biblio}

\end{document}